\documentclass[useAMS,usenatbib]{mn2e}

\usepackage{graphicx}
\newcommand{\NH}{N_\mathrm{H}}

\title[X-ray models for tori in AGNs]{A direct comparison of X-ray spectral models for tori in active galactic nuclei}
\author[Yuan Liu and Xiaobo Li]{Yuan Liu$^{1}$\thanks{E-mail:
liuyuan@ihep.ac.cn} and Xiaobo Li$^{1}$\thanks{E-mail: lixb@ihep.ac.cn}\\
$^{1}$Key Laboratory of Particle Astrophysics, Institute of High
Energy Physics \\ Chinese Academy of Sciences, P.O.Box 918-3,
Beijing 100049, China\\
}
\begin{document}

\date{}

\pagerange{\pageref{firstpage}--\pageref{lastpage}} \pubyear{2002}

\maketitle

\label{firstpage}

\begin{abstract}
Several X-ray spectral models for tori in active galactic nuclei
(AGNs) are available to constrain the properties of tori; however,
the accuracy of these models has not been verified. We recently
construct a code for the torus using Geant4, which can easily handle
different geometries \citep{b9}. Thus, we adopt the same
assumptions as \citet[][hereafter MY09]{b10} and  \citet[][hereafter
BN11]{b3} and try to reproduce their spectra. As a result, we can
reproduce well the reflection spectra and the strength of the Fe K$\alpha$ line of MY09, for  both $\NH=10^{24}$
and $10^{25}$ cm$^{-2}$. However, we cannot produce the strong
reflection component of BN11 in the low-energy band. The origin of this
component is the reflection from the visible inner wall of the torus, and it
 should be very weak in the edge-on directions under the geometry
of BN11. Therefore, the behaviour of the reflection spectra in BN11 is
not consistent with their geometry. The strength of the Fe K$\alpha$ line of BN11 is also different from our results and the analytical result in the optically thin case. The limitation of the spectral model will bias the parameters from X-ray spectral fitting.

\end{abstract}


\section{Introduction}

Under the unification scheme of active galactic nuclei (AGNs), a
toroidal structure referred as ``the torus" provides anisotropic
obscuration and can explain the diversity of optical and X-ray
spectra of AGNs \citep{a93}. The torus absorbs the intrinsic X-ray
spectra of AGNs (usually modelled as an absorbed power law) and also
scatters the X-ray photons to produce a Compton hump at $\sim$20
keV. The abundant and high quality data from recent X-ray satellites
enable us to investigate the X-ray properties of AGNs in
unprecedented detail. The structure of the tori in AGNs can be
constrained if an X-ray spectral model of the tori is specified.
Compared with the models for the disk geometry, e.g. \texttt{pexrav}
and \texttt{pexmon} \citep{m95,n07}, several more physical and
realistic models have been recently constructed to model the X-ray
spectrum from the torus, which adopt a toroidal structure and
self-consistently include fluorescent lines (Ikeda et
al 2009, hereafter IK09; Murphy \& Yaqoob 2009, hereafter MY09; Brightman \& Nandra 2011, hereafter BN11). These models have been applied to
individual AGNs to derive the covering factor and column density of
the sources \citep{b12,b13,b1,b7}. With a survey sample, these
models are helpful for finding Compton thick AGNs and determining how
the tori evolve with the properties of AGNs, such as the correlation
between the covering factor and luminosity \citep{b2,b5,b4,b6,b11}.
In spite of the success of the application of such models, the
accuracy and validity of such models have not been independently
verified. The parameters from these models will be biased if there
are some limitations or errors in these models. We recently
constructed an X-ray spectral model for the tori in AGNs using Geant4,
which can handle the smooth and clumpy tori by the same code. Thus,
if we adopt the same assumptions used by MY09, BN11 and IK09 (e.g.
geometries, cross sections and element abundances), our Geant4 code
should reproduce their result, in principle. In Section 2, we present
the quantitative comparison with the public models in MY09 and BN11.
Because the results of IK09 are not public, we just use the general
trend presented in IK09 as a reference. In Section 3, we discuss the
difference found in the comparison, and we show our main
conclusions in Section 4.

\section[]{Simulations for comparison}
The details of our simulation method are presented in \cite{b9}. We
have included several physical processes in our code, e.g.
photoelectric effect, Compton scattering, Rayleigh scattering,
$\gamma$ conversion, fluorescent lines and Auger process. Similar
physical processes have been considered in the simulations of MY09
and BN11. Thus, it is convenient to modify our code to simulate their cases.
Our code and the simulations by MY09 and BN11 have included
multi-scatterings, e.g. Figure 7 in \cite{b9}. We then use the same
assumptions adopted by BN11 and MY09 respectively to try to
reproduce their results\footnote[1]{The model file of MY09 is
downloaded from
http://mytorus.com/model-files-mytorus-downloads.html The model file of BN11
is downloaded from
http://www.mpe.mpg.de/$\sim$mbright/data/torus1006.fits}. The incident flux is $5\times10^8$ photons (1-500 keV) for all simulations presented in this paper.

\subsection{Results under the assumptions of MY09}

\begin{figure}
 \center
 \includegraphics[width=0.8\linewidth]{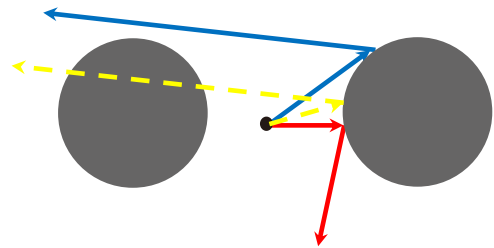}
  \caption{Geometry of MY09. The red and blue arrows indicate the possible trajectories of low-energy scattered photons that can escape to face-on and edge-on directions, respectively. The yellow arrows indicate the trajectories of low-energy scattered photons that will be absorbed by the near side of the torus for edge-on directions.}
\label{MY_geo}
\end{figure}

The geometry of MY09 is shown in Figure \ref{MY_geo}. The half-opening
angle of the torus is 60$^\circ$.

\begin{figure*}
  \center
   \includegraphics[width=0.8\linewidth]{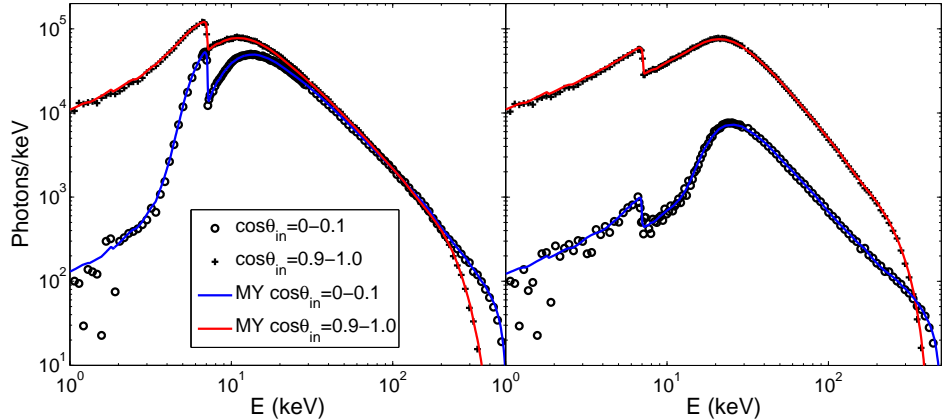}
  \caption{Spectra under MY09's assumptions with $\NH=10^{24}$ (left) and $10^{25}$ cm$^{-2}$ (right).
  Only scattered continua are shown for clarity. The scattered fluorescent photons are treated as the portion of emission lines and not shown here.}
  \label{MY_com}
\end{figure*}

Using the same geometry, cross sections, element abundances and
incident spectra (a single power law with the photon index
$\Gamma=1.8$), we can well reproduce MY09's continua, for both
$\NH=10^{24}$ and $10^{25}$ cm$^{-2}$ (Figure \ref{MY_com},
where $\theta_{\mathrm{in}}$ is the inclination angle relative to an
observer). The direct component is not shown for clarity. Since the
shape of the scattered component is determined by both the
scattering and absorption processes (e.g. Fe K absorption edge at 7
keV), these results verify the accuracy of MY09's simulations.

We have further compared the equivalent width (EW) of the Fe K$\alpha$ line of our simulations with that in  Figure 8 of \cite{my11}, which is the erratum of MY09. The photon index $\Gamma$ is 1.9 in these simulations. For $\NH=10^{24}$ and $10^{25}$ cm$^{-2}$, the difference of EWs between our simulations and MY09's model is smaller than 1\% for most of the directions. The maximum deviation is $\sim$2\% for $\NH=10^{24}$ cm$^{-2}$ and $\cos\theta_{\mathrm{in}}=0.3-0.4$. This small deviation could be due to the different simulation method adopted by MY09. The accuracy of this model is sufficient, since the statistical error of EW(Fe K$\alpha$) in observed X-ray spectra is larger than 10\% in most cases \citep{L10,s11}.

\subsection{Results under the assumptions of BN11}
\begin{figure}
\center
\includegraphics[width=0.5\linewidth]{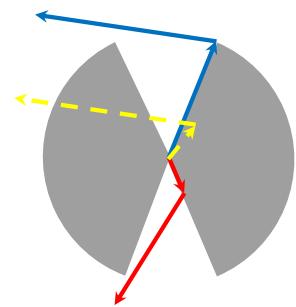}
  \caption{Geometry of BN11. The red and blue arrows indicate the possible trajectories of low-energy scattered photons that can escape to face-on and edge-on directions, respectively. The yellow arrows indicate the trajectories of low-energy scattered photons that will be absorbed by the near side of the torus for edge-on directions.}
  \label{BN_geo}
\end{figure}

The geometry of BN11 is shown in Figure \ref{BN_geo}. The half-opening
angle of the torus is fixed at 60$^\circ$.

\begin{figure*}
   \center
   \includegraphics[width=0.8\linewidth]{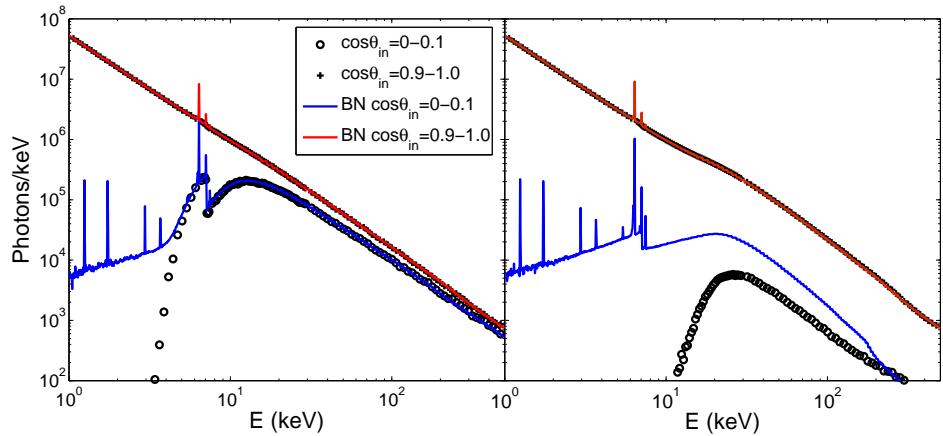}
  \caption{Spectra under BN11's assumptions with $\NH=10^{24}$ (left) and $10^{25}$ cm$^{-2}$ (right).
  The total spectra (scattered$+$direct) are shown, but the fluorescent lines (including the Compton shoulders) in our simulation
   are not shown for clarity. The half-opening angle is assumed to be 60$^\circ$.}
  \label{BN_com}
\end{figure*}

The inner radius of the torus is assumed to be zero in BN11's model.
The comparison between the BN11's model and our results is shown in
Figure \ref{BN_com} ($\Gamma$=1.8). The total spectra are shown, i.e. direct$+$
scattered components, since BN11's model only provides the total
spectra. The fluorescent lines (including the Compton shoulders) in the spectra of our simulations are not shown for clarity, because it
is not easy to directly compare the strength of such a
quasi-$\delta$ function in the figure.

Although the continua of the face-on directions are consistent
with each other, the spectra of the edge-on directions, especially
for low-energy band, are significantly different. It seems there is
an ``additional'' component at the low-energy band in the
simulations of BN11.

We have further compared  EW(Fe K$\alpha$) of our simulations with that in Figure 3 of BN11. The photon index $\Gamma$ is 2.0 in these simulations. For the face-on direction ($\theta_{\mathrm{tor}}=60^\circ$ and $\theta_{\mathrm{in}}=0-37^\circ$), EW(Fe K$\alpha$) of BN11 is lower than our simulation results by 30\% and 35\% for $\NH=10^{24}$ and $10^{25}$ cm$^{-2}$, respectively. For the edge-on direction ($\theta_{\mathrm{tor}}=60^\circ$ and $\theta_{\mathrm{in}}=78-90^\circ$), their EW(Fe K$\alpha$) is higher than our results by 60\% for $\NH=10^{24}$ cm$^{-2}$. Due to the large deviation of the continuum of the edge-on direction for $\NH=10^{25}$ cm$^{-2}$, we have not compared the EW(Fe K$\alpha$) in this case.

We explore these discrepancies in detail in the next
section.

\section{Discussion}
Under the $\NH$ discussed in Figure \ref{MY_com} and \ref{BN_com},
the direct component is highly absorbed in the low-energy band.
Thus, the low-energy spectra are dominated by the reflection from
the visible inner wall of the torus for a given inclination angle,
which is referred to as the `reflection component 2'  in IK09 (the
Figure 2 in IK09 illustrates its geometry and origin).

The strength of this component depends on the geometry of the torus,
i.e. the location and shape of the surface, and should also depend
on the inclination angles. Therefore, to judge whether the results
of BN11's model in Figure \ref{BN_com} are reasonable, we
investigate the variation of this component with inclination angles
in different models.

\begin{figure}
  \center
  \includegraphics[width=0.8\linewidth]{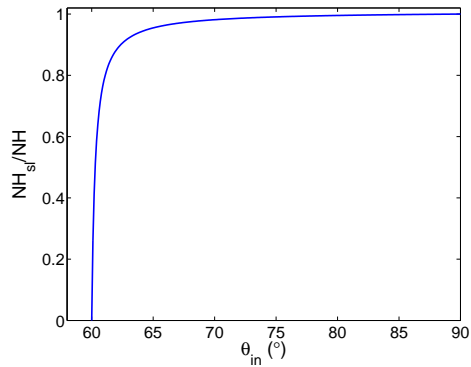}
  \caption{Ratio between the column density along the line of sight and the maximum $\NH$ in IK09 (half-opening angle=$60^\circ$ and $r_{\mathrm{in}}/r_{\mathrm{out}}=0.01$). It is almost constant except for the angles near the edge of the torus.}
  \label{IK_NH}
\end{figure}

Because $r_{\mathrm{in}}/r_{\mathrm{out}}=0.01$ adopted in the
simulations of IK09 is small, IK09's geometry is actually very
similar to that of BN11, i.e. the $\NH$ distribution is almost
constant for different inclination angles. We plot the distribution
of $\NH$ using equation (3) in IK09 to show this (Figure
\ref{IK_NH}).

As shown in the Figure 9 in IK09, the `reflection component 2'
significantly decreases with increasing inclination angles, i.e. it
is very weak in the edge-on directions.

\begin{figure}
  \center
  \includegraphics[width=0.8\linewidth]{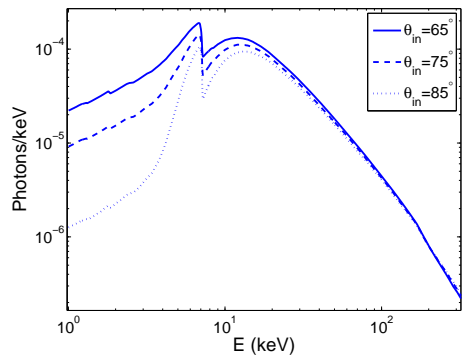}
  \caption{The scattered component is significantly suppressed in the edge-on direction under MY09's geometry.}
  \label{MY3in}
\end{figure}

A similar trend is also observed in MY09's model (Figure
\ref{MY3in}), though the variation is not as dramatic as that in the
Figure 9 in IK09. We will explain this later. The spectrum at 1 keV
with inclination angle $\theta_{\mathrm{in}}=65^\circ$ is higher
than that with $\theta_{\mathrm{in}}=85^\circ$ by more than one order.

\begin{figure}
  \center
  \includegraphics[width=0.8\linewidth]{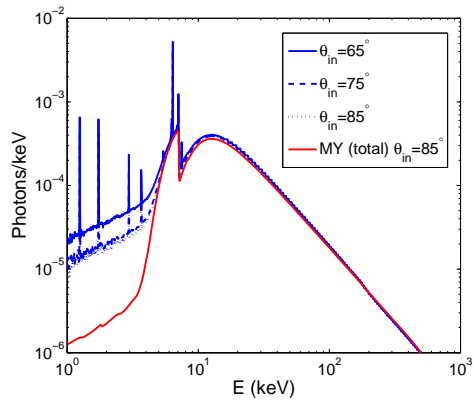}
  \caption{The scattered component only weakly depends on the inclination angels in BN11's model. The strength of the low-energy component in edge-on direction is much higher than that produced by MY09's model (scattered and direction components are added).}
  \label{BN3in}
\end{figure}

However, the strength of this low-energy reflection component in
BN11's model only weakly depends on the inclination angles (Figure
\ref{BN3in}). The spectrum at 1 keV with
$\theta_{\mathrm{in}}=65^\circ$ is only higher than that with
$\theta_{\mathrm{in}}=85^\circ$ by a factor of two.

The strength of this low-energy component depends on the visibility
of the inner wall of the torus at different inclination angles.
Under the geometry of MY09, since the central part of the torus is
empty, the whole inner surface of the torus is directly illuminated
by the central source. As a result, a considerable part of the inner
wall is visible when the inclination angle is slightly larger than
the half-opening angle of the torus (the blue arrows in Figure
\ref{MY_geo}); even for the edge-on case, the rim of the inner wall
is still visible but the majority of reflection in low-energy band
is absorbed by the near side of the torus (the blue and yellow
arrows in Figure \ref{MY_geo}). This can explain the trend observed
in Figure \ref{MY3in}.

\begin{figure}
 \center
 \includegraphics[width=1.0\linewidth]{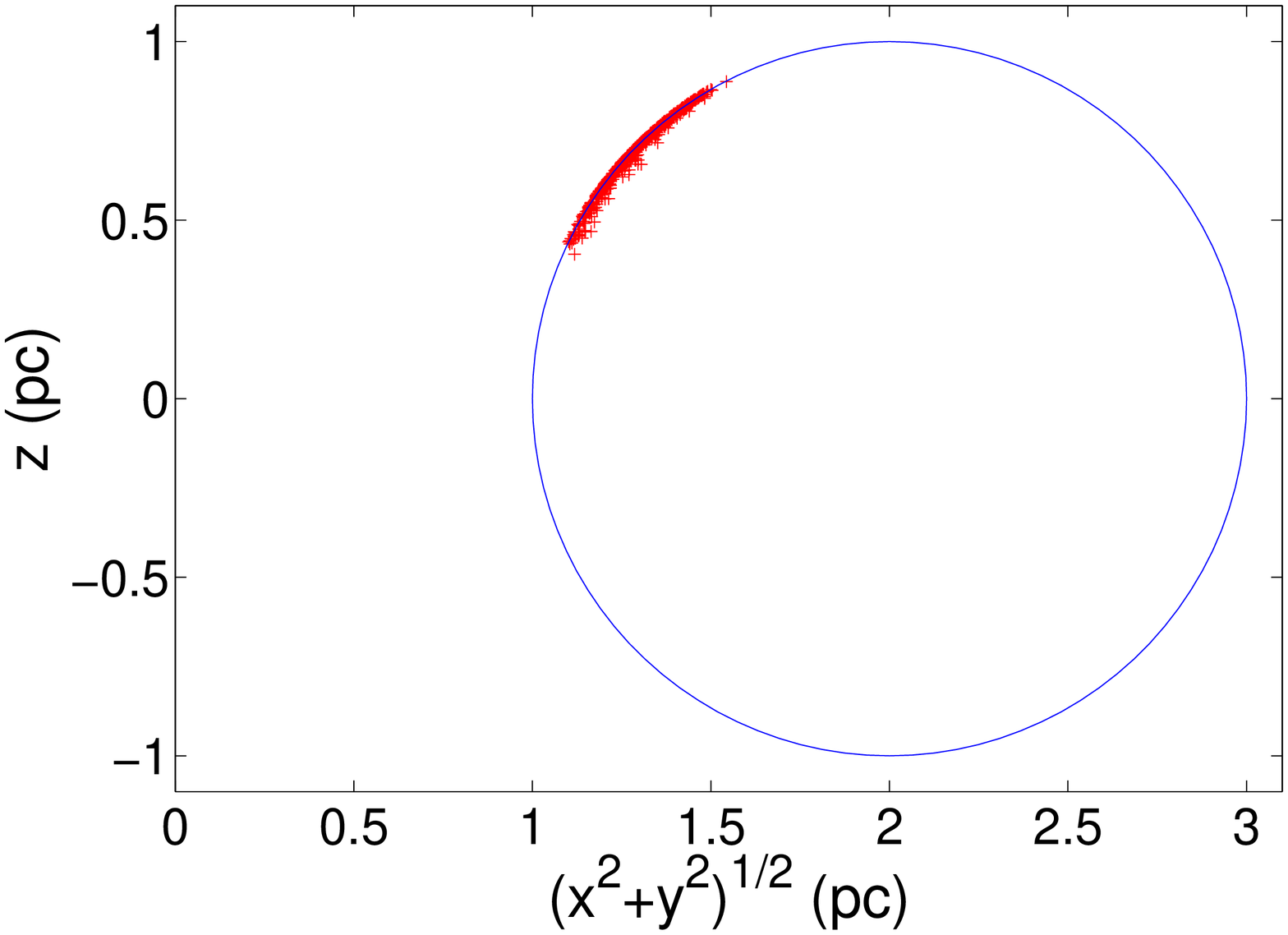}
 \includegraphics[width=1.0\linewidth]{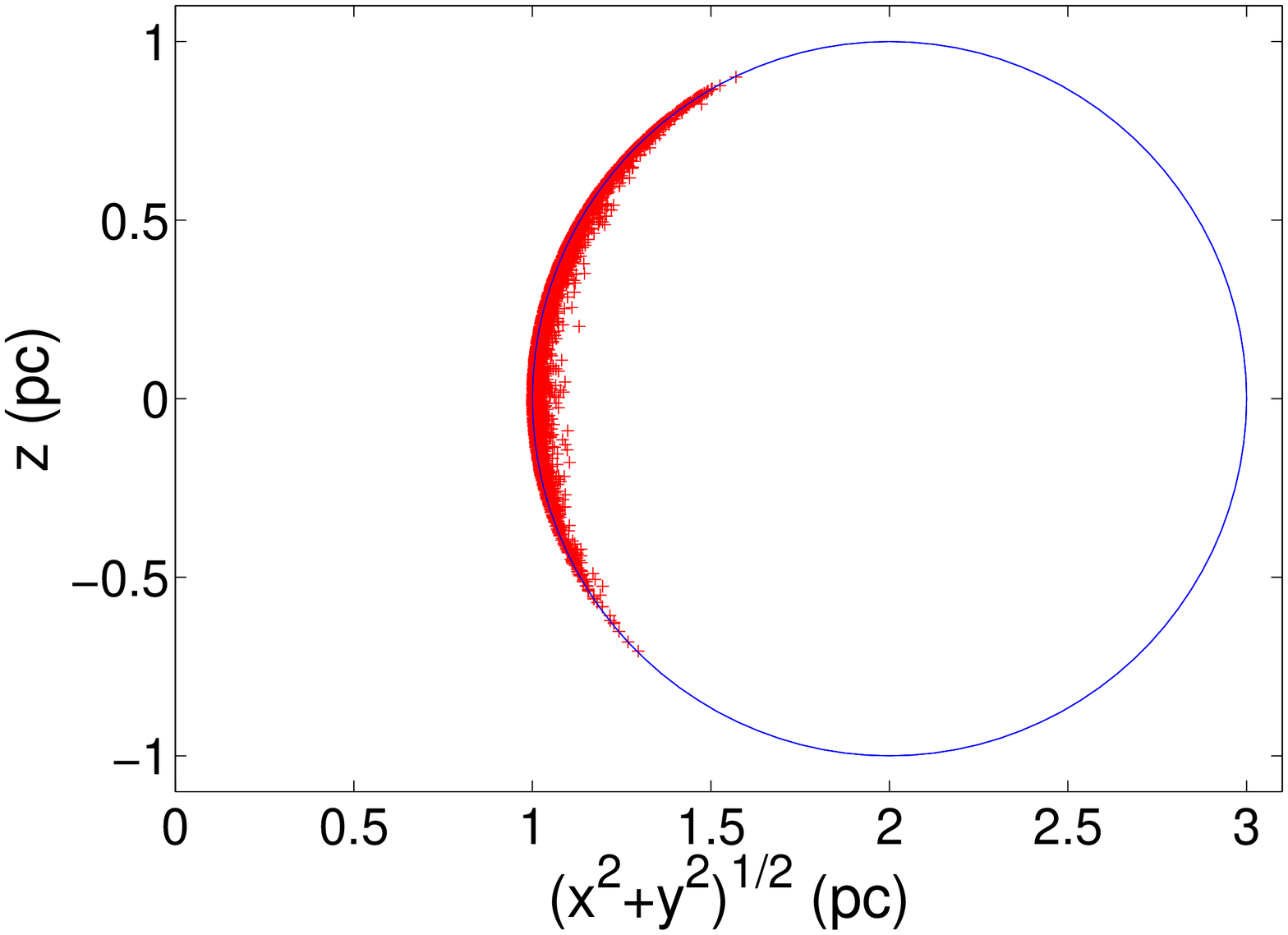}
  \caption{Positions of the scatterings of the photons (1-2 keV) escaped to $\cos\theta_{\mathrm{in}}=0-0.2$ (top) and $\cos\theta_{\mathrm{in}}=0.4-0.5$ (bottom) under MY09's geometry. The 3-D positions are projected onto a plane to show the distribution more clearly. The blue line shows the boundary of the torus and the central X-ray source is located at the origin.}
  \label{MY_pos}
\end{figure}

To further support the above explanation, in Figure \ref{MY_pos} we also plot the positions
of the scatterings of the observed photons in 1-2 keV, i.e. the
photons have experienced scatterings and finally escaped to the
observer. The distributions of
the positions of the scatterings for two directions are shown. The scatterings
of low-energy photons can only occur at the skin of the torus;
otherwise, they will be absorbed in the body of the torus. As the
inclination angle moves to the edge-on direction, the visible part
also moves to the rim. If the scattered photons intend to escape
from the edge-on direction, the scatterings should occur at the rim
of the torus; otherwise, they will be absorbed by the near side of
the torus.

\begin{figure}
 \center
 \includegraphics[width=0.8\linewidth,height=4.8cm]{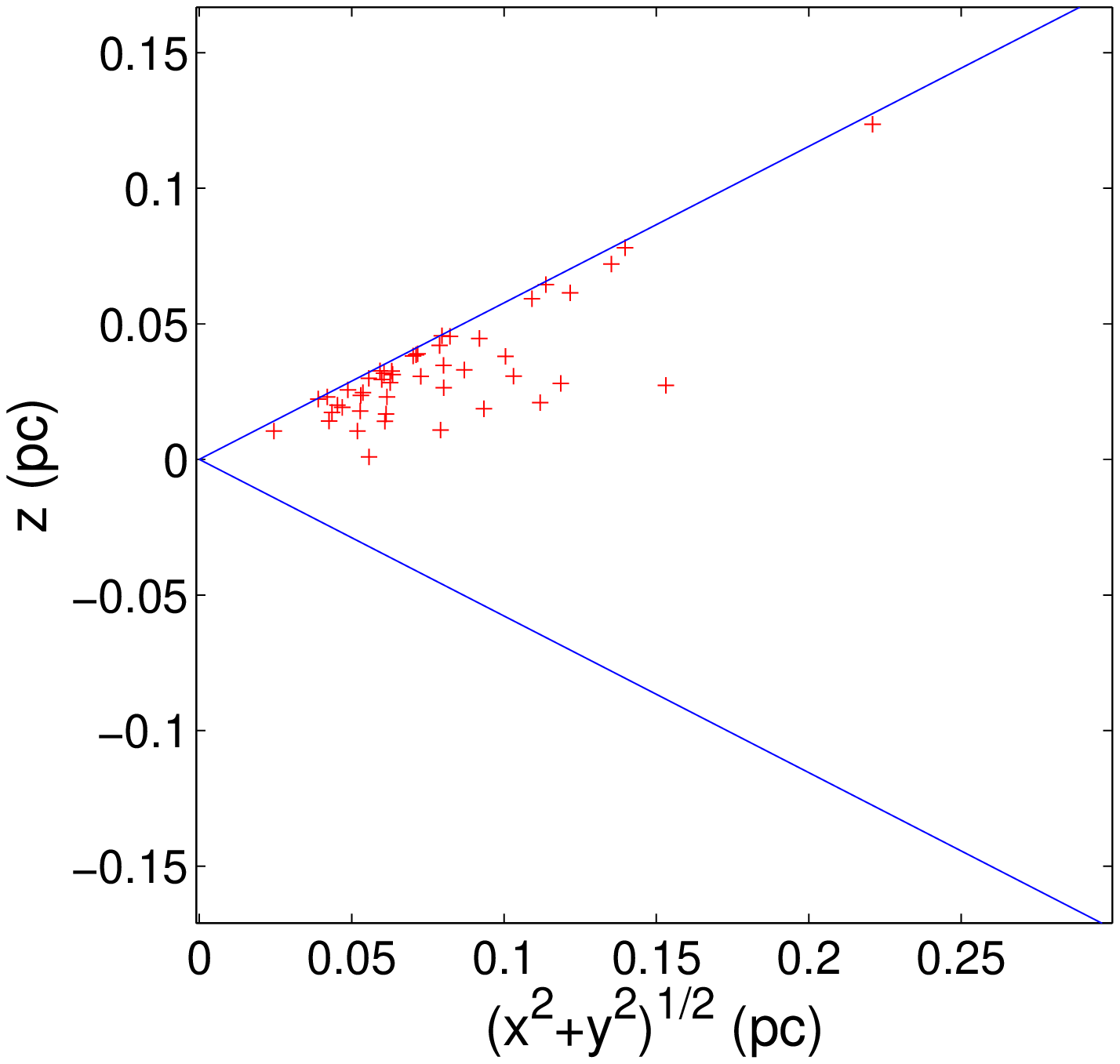}
 \includegraphics[width=0.8\linewidth,height=4.8cm]{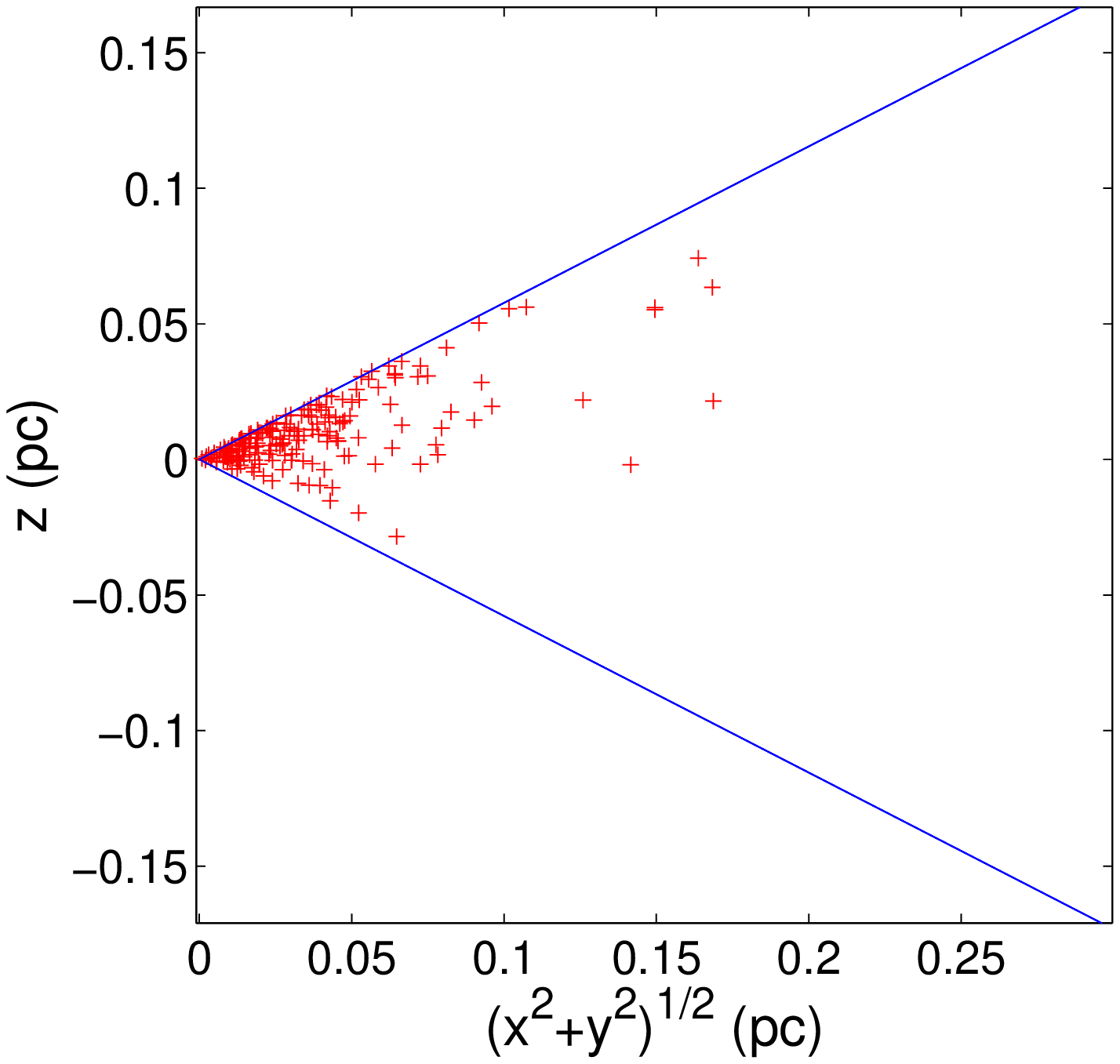}
  \caption{Positions of the scatterings of the photons (1-2 keV) escaped to $\cos\theta_{\mathrm{in}}=0-0.49$ (top) and $\cos\theta_{\mathrm{in}}=0.49-0.5$ (bottom) under BN11's geometry.
  The intervals of $\cos\theta_{\mathrm{in}}$ are different from that in Figure \ref{MY_pos}, since the number of the
  scattered photons under BN's geometry rapidly decreases when the inclination angle is larger than
  the half-opening angle of the torus by a few degrees.
  The 3-D positions are projected onto a plane to show the distribution more clearly. The inner radius of the torus is 0 pc and the outer radius is 2 pc (not shown). The blue line shows the boundary of the torus and the central X-ray source is located at the origin.}
  \label{BN_pos}
\end{figure}

For the BN11 geometry, the torus extends to the centre and the
column densities are the same for different inclination angles. As a
result, if the low-energy photons are scattered and can escape to
the observer, they can only be scattered very near to the centre;
otherwise, they will be absorbed before reaching a large radius.
Therefore, the reflection component is only visible when the
inclination is slightly larger than the half-opening angle (a few
degrees). The possible trajectory of the scattered photons is
indicated by the arrows in Figure \ref{BN_geo}. For the edge-on
case, the scattered region is obscured by the near side of the
torus. If the scattered photons intend to escape to the edge-on
direction, they should be scattered at a large radius and follow
the way indicated by the blue arrows in Figure \ref{BN_geo}.
However, such photons should be rare, since most of them will be
absorbed before reaching the large radius. This is also the reason
for the significant decrease found in the Figure 9 in IK09. We plot
the positions of the scatterings in Figure \ref{BN_pos}, which are
indeed concentrated in the central part of the torus. The region is
very small compared with the outer radius (2 pc) of the torus.
Therefore, this reflection component in the edge-on direction should
be very weak under the geometry of BN11 and IK09, as shown by our
simulations in Figure \ref{BN_com}.

As mentioned in Section 2.2, the EW(Fe K$\alpha$) of BN11 is also different from our simulations. For optically thin case, EW(Fe K$\alpha$) can be analytically calculated, e.g. by equation (5) in MY09, and nearly isotropic. Thus, we take this analytical result as the benchmark test of BN11's model and our simulations. Under the assumption in Figure 3 of BN11, the analytical EW(Fe K$\alpha$) is 3.7 eV for $\Gamma=2.0$, $\theta_{\mathrm{tor}}=60^\circ$, and $\NH=10^{22}$ cm$^{-2}$. The EW(Fe K$\alpha$) of our simulation under the same assumption is well consistent with 3.7 eV (at better than 1\% level). However, the EW of BN11's model is about 4.4 eV for the same case. Since their continua at 6.4 keV are consistent with our results at 1\% level even for the edge-on case of $\NH=10^{24}$ cm$^{-2}$, there should be some problems in the transportation of Fe K$\alpha$ photons, which might  be related to the problem inducing the overestimate of the reflection component.

\section{Conclusions}

With the code constructed using Geant4, we can reproduce well the continua and the strength of Fe K$\alpha$ line of MY09's model. However, the reflection component in the
low-energy band is much lower than that of BN11's model. We have
discussed the origin of this reflection component and shown that the
scattered region is concentrated in the centre and invisible in the
edge-on directions under the BN11's geometry. Therefore, it seems the strength of the reflection component is overestimated in BN11 for the edge-on
directions. The strength of Fe K$\alpha$ line of BN11's model is also different from our results and the analytical result in the optically thin case, which is likely to be due to the problem in the transportation of Fe K$\alpha$ photons. The accuracy of the model is crucial to any conclusions from the spectral fitting.

\section*{Acknowledgments}

This work is supported by the National Natural Science Foundation of
China under grant Nos. 11103019, 11303027, 11133002, and 11103022.

\bsp

\label{lastpage}

\end{document}